\documentclass[12pt]{article}

\usepackage{amsmath}
\usepackage{amsfonts}
\usepackage{graphicx}
\usepackage{caption}
\usepackage{subcaption}

\usepackage{authblk}
\usepackage{amssymb}

\usepackage{epstopdf}
\usepackage{epsfig}
\usepackage{hyperref}

\usepackage{color}

\date{}

\begin{document}

\title{Axial perturbations of the scalarized Einstein-Gauss-Bonnet black holes}

\author[1]{Jose Luis Bl\'azquez-Salcedo \thanks{\href{mailto:jose.blazquez.salcedo@uni-oldenburg.de}{jose.blazquez.salcedo@uni-oldenburg.de}}}

\author[2,3]{Daniela D. Doneva
\thanks{\href{mailto:daniela.doneva@uni-tuebingen.de }{daniela.doneva@uni-tuebingen.de }}}

\author[1]{Sarah Kahlen\thanks{\href{mailto:sarah.kahlen1@uni-oldenburg.de}{sarah.kahlen1@uni-oldenburg.de}}}

\author[1]{Jutta Kunz \thanks{\href{mailto:jutta.kunz@uni-oldenburg.de}{jutta.kunz@uni-oldenburg.de}}}

\author[1,4]{Petya Nedkova \thanks{\href{mailto:pnedkova@phys.uni-sofia.bg}{pnedkova@phys.uni-sofia.bg}}}

\author[2,4,5]{Stoytcho S. Yazadjiev \thanks{\href{mailto:yazad@phys.uni-sofia.bg}{yazad@phys.uni-sofia.bg}}}

\affil[1]{Institute of Physics, Carl von Ossietzky University of Oldenburg, 26111
	Oldenburg, Germany}
\affil[2]{Theoretical Astrophysics, Eberhard Karls University of T\"ubingen, 72076 T\"ubingen, Germany}
\affil[3]{INRNE - Bulgarian Academy of Sciences, 1784  Sofia, Bulgaria}
\affil[4]{Department of Theoretical Physics, Faculty of Physics, Sofia University, 1164 Sofia, Bulgaria}
\affil[5]{Institute of Mathematics and Informatics, Bulgarian Academy of Sciences, Acad. G. Bonchev Street 8, 1113 Sofia, Bulgaria}

\maketitle

\begin{abstract}
We study the axial perturbations of spontaneously scalarized black holes
in Einstein-Gauss-Bonnet (EGB) theories.
We consider the nodeless solutions of the fundamental branch
of the model studied in \cite{Doneva:2017bvd},
which possesses a region of radially stable configurations,
as shown in \cite{Blazquez-Salcedo:2018jnn}.
Here we show that
almost all of the radially stable black holes
are also stable under axial perturbations.
When the axial potential is no longer strictly positive,
we make use of the S-deformation method to show stability.
As for the radial perturbations, hyperbolicity is lost
below a certain critical horizon size for a fixed coupling constant.
In the stable region, we determine the spectrum of the quasinormal modes
by time evolution and by solving the associated time-independent
eigenvalue problem.
\end{abstract}
%\pacs{}

\section{Introduction}
Alternative theories of gravity are attracting more and more attention
lately, especially in the light of the exciting prospects to use
the gravitational wave observations to test regimes and predictions
that can not be addressed with the present capabilities
of the electromagnetic observations \cite{TheLIGOScientific:2016src,Abbott:2018lct,LIGOScientific:2019fpa,Will:2014kxa,Berti:2015itd,Barack:2018yly}.
Particularly interesting are classes of alternative theories
that are indistinguishable from general relativity (GR)
in the weak field regime, but lead to very interesting
nonlinear effects for large spacetime curvatures
\cite{Will:2005va,Capozziello:2010zz,Berti:2018cxi,Berti:2018vdi}. 
The reason is that in this way, all the weak field tests of GR,
that are often performed with an amazing accuracy,
are automatically satisfied, while we are still left with the possibility
to explore the very interesting and not yet well constrained
strong field regime of gravity via astrophysical observations.

Historically, the first theories where such nonlinear strong field effects
were examined are the scalar-tensor theories of gravity
\cite{Brans:1961sx,Damour:1992we},
where it was found that for a certain range of central energy densities
and coupling parameters,
a neutron star can develop a nonlinear scalar field
sourced by the trace of the energy-momentum tensor of its matter
\cite{Damour:1993hw,Damour:1996ke}.
Even more interestingly, such
matter-induced scalarized neutron stars are energetically
more favorable than the GR solutions with vanishing scalar field,
and thus they are the ones that would be realized in nature.

A natural question to ask is whether the effect of spontaneous scalarization
can be observed for black holes as well.
Very recently, it was discovered that the spacetime curvature itself
can act as a source for the scalar field
and thus lead to scalarized black holes
in the framework of the Gauss-Bonnet theories \cite{Doneva:2017bvd,Silva:2017uqg,Antoniou:2017acq,Antoniou:2017hxj}.
These results for curvature-induced spontaneously scalarized
static spherically symmetric black holes
were extended to the case of charged black holes
\cite{Doneva:2018rou},
black holes with a massive scalar field \cite{Macedo:2019sem,Doneva:2019vuh}
and black holes with a cosmological constant \cite{Bakopoulos:2018nui,Brihaye:2019gla}.
Moreover, rotating black holes and excited black holes were
studied \cite{Cunha:2019dwb,Collodel:2019kkx}.
It was also observed that the spontaneous scalarization of
black holes can be charge-induced
\cite{Herdeiro:2018wub,Blazquez-Salcedo:2020nhs}
(see also \cite{Stefanov:2007eq,Doneva:2010ke}
for earlier work in this connection). The stability of the scalarized solutions was discussed in \cite{Blazquez-Salcedo:2018jnn,Minamitsuji:2018xde,Silva:2018qhn,Macedo:2019sem}.

In \cite{Blazquez-Salcedo:2018jnn}
it was demonstrated that the Schwarzschild solution loses stability
at the point where the first scalarized EGB branch of black holes appears,
which is characterized by a scalar field without nodes.
As the horizon radius decreases, a sequence of further branches
of scalarized black holes bifurcates from the Schwarzschild solution,
where each branch can be labeled by the number of zeros of the scalar field.
The most important question to be answered is thus
whether these scalarized branches are stable or not.
This question was first addressed in \cite{Blazquez-Salcedo:2018jnn},
where it turned out that all the additional branches
with a scalar field with one or more nodes are unstable
under radial perturbations.
In fact, depending on the choice of the coupling function,
only the first fundamental scalarized branch can be stable.
The different terms of the coupling functions
that can stabilize the fundamental branch
were further explored in \cite{Silva:2018qhn}.
Stabilization by a quartic self-interacting scalar field potential
is also possible \cite{Minamitsuji:2018xde,Macedo:2019sem}.
Later, studies showed that curvature-induced spontaneous scalarization
is also possible for a generalized version of the Gauss-Bonnet theories -- the Horndeski theories
\cite{Minamitsuji:2019iwp,Horndeski:1974wa}.

In view of the huge advance of the gravitational wave observations,
it is very important to study the gravitational wave signatures
of such scalarized objects.
As a first step in this direction, we consider in the present
paper their axial quasinormal modes (QNMs),
since QNMs describe the ringdown after a merger.
The general set of scalar and metric perturbations
can be decomposed into parity-even (polar) and parity-odd (axial) perturbations.
The radial perturbations, studied before for scalarized EGB black holes in
\cite{Blazquez-Salcedo:2018jnn,Silva:2018qhn,Macedo:2019sem},
represent a special set of polar perturbations.
In contrast to the polar modes,
the axial modes do not involve the perturbations of the scalar field,
leading therefore to a simpler spectrum
as, for instance, seen in previous work on polar and axial QNMs of
scalarized black holes in dilatonic EGB theory
\cite{Pani:2009wy,Blazquez-Salcedo:2016enn,Blazquez-Salcedo:2017txk,Blazquez-Salcedo:2018pxo,Konoplya:2019hml,Zinhailo:2019rwd}.

Here we will focus not only on the QNM frequencies,
but also on the question whether the scalarized black holes are stable
under such perturbations.
The reason this question is interesting
is because even if we choose a proper coupling function,
stability needs not to hold on the full domain of existence
of the scalarized black holes.
In fact, as shown in \cite{Blazquez-Salcedo:2018jnn},
the radial perturbation equations lose hyperbolicity
for small values of the horizon radius, $r_{\rm H} \le r_{\rm S1}$,
and fixed coupling constant.
Here we will show that hyperbolicity of the
axial perturbation equations will be lost slightly earlier,
at $r_{\rm S2} > r_{\rm S1}$.
To show stability with respect to axial perturbations
when the potential is no longer strictly positive,
we will make use of the S-deformation method
\cite{Kimura:2017uor,Kimura:2018whv,Blazquez-Salcedo:2020nhs}.

In Section II we will present the EGB theory employed, recalling
the conditions on the coupling function to obtain spontaneously
scalarized black holes. The axial perturbations are discussed
in Section III, where the master equation is derived and the
proper set of boundary conditions is stated.
Section IV contains our results.
First, the loss of hyperbolicity is discussed,
then the S-deformation method is employed to show stability
when the potential is no longer strictly positive,
and finally the axial QNMs are presented for the fundamental
$l=2$ mode. The modes are obtained in two ways,
by the method of time evolution and
by solving the time-independent eigenvalue problem.
The paper ends with conclusions.

\section{Gauss-Bonnet gravity}

The action in Gauss-Bonnet gravity has the following form:
\begin{eqnarray}
S=&&\frac{1}{16\pi}\int d^4x \sqrt{-g}
\Big[R - 2\nabla_\mu \varphi \nabla^\mu \varphi
+ \lambda^2 f(\varphi){\cal R}^2_{GB} \Big] \ ,\label{eq:quadratic}
\end{eqnarray}
where $R$ is the Ricci scalar with respect to the spacetime metric $g_{\mu\nu}$,
$\varphi$ is the scalar field, $f(\varphi)$ is the coupling function,
that depends on $\varphi$ only,
and $\lambda$ is the Gauss-Bonnet coupling constant
that has dimension of $length$.
The Gauss-Bonnet invariant ${\cal R}^2_{GB}$ is defined as
${\cal R}^2_{GB}=R^2 - 4 R_{\mu\nu} R^{\mu\nu}
+ R_{\mu\nu\alpha\beta}R^{\mu\nu\alpha\beta}$,
where $R_{\mu\nu}$ is the Ricci tensor
and $R_{\mu\nu\alpha\beta}$ is the Riemann tensor.

The field equations derived after varying the action above are given by
\begin{eqnarray}\label{FE1}
&&R_{\mu\nu}- \frac{1}{2}R g_{\mu\nu} + \Gamma_{\mu\nu}= 2\nabla_\mu\varphi\nabla_\nu\varphi -  g_{\mu\nu} \nabla_\alpha\varphi \nabla^\alpha\varphi \ ,\\
&&\nabla_\alpha\nabla^\alpha\varphi=  -  \frac{\lambda^2}{4} \frac{df(\varphi)}{d\varphi} {\cal R}^2_{GB} \ , \label{FE2}
\end{eqnarray}
where $\Gamma_{\mu\nu}$ is defined as
\begin{eqnarray}
\Gamma_{\mu\nu}&=& - R(\nabla_\mu\Psi_{\nu} + \nabla_\nu\Psi_{\mu} ) - 4\nabla^\alpha\Psi_{\alpha}\left(R_{\mu\nu} - \frac{1}{2}R g_{\mu\nu}\right) +
4R_{\mu\alpha}\nabla^\alpha\Psi_{\nu} + 4R_{\nu\alpha}\nabla^\alpha\Psi_{\mu} \nonumber \\
&& - 4 g_{\mu\nu} R^{\alpha\beta}\nabla_\alpha\Psi_{\beta}
+ \,  4 R^{\beta}_{\;\mu\alpha\nu}\nabla^\alpha\Psi_{\beta} \ ,
\end{eqnarray}
and
\begin{eqnarray}
\Psi_{\mu}= \lambda^2 \frac{df(\varphi)}{d\varphi}\nabla_\mu\varphi \ .
\end{eqnarray}

In the present paper, for simplicity we will work with
zero scalar field potential, $V(\varphi) = 0$,
and assume that the cosmological value of the scalar field is zero,
$ \varphi_{\infty}=0$.
We will be interested in spontaneously scalarized black hole solutions.
Thus, the coupling function $f(\varphi)$ should satisfy
the conditions
$\frac{df}{d\varphi}(0)=0$ and $b^2=\frac{d^2f}{d\varphi^2}(0)>0$,
where we can assume that $b=1$ without loss of generality.

The specific form of $f(\varphi)$
we will use in the present paper is the following:
\begin{equation} \label{eq:coupling_function}
f(\varphi)=  \frac{1}{12} \left(1- e^{-6\varphi^2}\right) \ .
\end{equation}
{The advantage of this choice is that for a certain region of the parameter space, we expect to have nicely behaving  scalarized solutions \cite{Doneva:2017bvd} that are as well stable against radial perturbations \cite{Blazquez-Salcedo:2018jnn}.}

We will be interested in spherically symmetric solutions
of the field equations and therefore the general ansatz
for the metric can be taken in the form
\begin{eqnarray}\label{eq:metric_BG}
ds^2= - e^{2\mu(r)}dt^2 + e^{2\nu(r)} dr^2
+ r^2 (d\theta^2 + \sin^2\theta d\phi^2 ) \ .
\end{eqnarray}
The reduced field equations after assuming the above metric ansatz
can be found in \cite{Doneva:2017bvd}.
Since the focus of the present paper is on the perturbations,
we will not give them explicitly here
and we refer the reader to \cite{Doneva:2017bvd} for further information.

\section{Axial Perturbations}

To obtain axial QNMs,
the perturbation of the background metric \eqref{eq:metric_BG}
can be written in the following form:
\begin{eqnarray}
&&ds^2=-e^{2\mu_0}dt^2+e^{2\nu_0}dr^2 +r^2\left[d\theta^2
+ \sin^2\theta (d\varphi+ k_1dt+  k_2dr+  k_3d\theta)^2\right] \ ,
\end{eqnarray}
where $\mu_0$ and $\nu_0$ are the background metric functions
depending on the $r$ coordinate only,
while the perturbations $k_1$, $k_2$ and $k_3$
are functions of the coordinates ${t, r, \theta}$.
Since the scalar field transforms as a true scalar
under reflections of the angular coordinates, its axial perturbation is zero.

{After perturbing the field equations \eqref{FE1}--\eqref{FE2},
it is possible to show that $k_1$ can be expressed algebraically
in terms of $k_2$ and $k_3$. After introducing the new variable }
\begin{equation}
\mathcal{Q}(t, r, \theta)= -r^2\sin^3\theta\, e^{\mu_0-\nu_0}\mathcal{P}(r)
\left(\partial_{\theta}k_{2}-\partial_{r}k_{3}\right) \ ,
\end{equation}
{it turns out that the perturbation equations can be reduced to a single second order equation for $\mathcal{Q}$:}
\begin{equation}\label{pert_eq}
r^4e^{-\mu_0-\nu_0}\,\mathcal{W}(r)\frac{\partial}{\partial r}\left[\frac{e^{\mu_0-\nu_0}}{r^2\mathcal{S}(r)}\frac{\partial\mathcal{Q}}{\partial r}\right]+\sin^3\theta\frac{\partial}{\partial \theta}\left[\frac{1}{\sin^3\theta}\frac{\partial\mathcal{Q}}{\partial \theta}\right] = \frac{r^2 e^{-2\mu_0}\,\mathcal{W}(r)}{\mathcal{P}(r)}\partial_{t}^2\mathcal{Q} \ ,
\end{equation}
where we have introduced the following auxiliary functions
of the background quantities:
\begin{eqnarray}
&&\mathcal{P}_0(r)=1-4\lambda^2\,\frac{df(\varphi_0)}{d\varphi_0}\,\mu_0'\,\varphi_0'\, e^{-2\nu_0} \ ,  \\[3mm]
&&\mathcal{W}_0(r)=1-4\lambda^2\frac{df(\varphi_0)}{d\varphi_0}\,\frac{\varphi_0'}{r}\, e^{-2\nu_0} \ ,  \\[3mm]
&&\mathcal{S}_0(r)=1-4\lambda^2\frac{d^2f(\varphi_0)}{d\varphi_0^2}(\varphi_0')^2\,e^{-2\nu_0} - 4\lambda^2\frac{df(\varphi_0)}{d\varphi_0}\,\varphi_0''\,e^{-2\nu} +  4\lambda^2\frac{df(\varphi_0)}{d\varphi_0}\,\nu_0'\,\varphi_0'\,e^{-2\nu_0}.
\end{eqnarray}
Here $\varphi_0$ is the background value of the scalar field, and
the derivative with respect to the radial coordinate $r$ is denoted by $()'$.

Since the background solutions are spherically symmetric,
the new variable can be separated in the form
$\mathcal{Q}=\hat Q(t,r)D(\theta)$.
Then, Eq.~($\ref{pert_eq}$) yields
\begin{eqnarray}
&& r^4e^{-\mu_0-\nu_0}\,\mathcal{W}_0\frac{\partial}{\partial r}\left[\frac{e^{\mu_0-\nu_0}}{r^2\mathcal{S}_0}\frac{\partial\hat Q}{\partial r}\right]-\alpha \hat Q=\frac{r^2 e^{-2\mu_0}\,\mathcal{W}_0}{\mathcal{P}_0}\partial_{t}^2\hat Q \ , \label{eq_sepa1} \\
&& \sin^3\theta\frac{d}{d \theta}\left[\frac{1}{\sin^3\theta}\frac{d D}{d \theta}\right]=-\alpha D \ , \label{eq_sepa2}
\end{eqnarray}
where $\alpha$ is a separation constant.
Eq.~($\ref{eq_sepa2}$) has a solution in terms of the Legendre polynomials
\begin{equation}
D(\theta)=\sin^3 \theta \frac{d}{d\theta}\frac{1}{\sin \theta}\frac{dP_{l}\left(\cos \theta \right)}{d\theta} \ ,
\end{equation}
where the separation constant takes the values $\alpha= (l-1)(l+2)$
with $l$ being a positive integer.
Eq.~($\ref{eq_sepa1}$) can be simplified by introducing
the Regge-Wheeler coordinate $\tilde{r}_{*}$,
defined by $\frac{\partial}{\partial r}
=e^{\nu_0-\mu_0}\frac{\partial}{\partial \tilde{r}_{*}}$,
and the master variable $\Psi(\tilde{r}_*,t)= \hat Q(r,t)/ r\mathcal{S}_0^{1/2}$. {In this way we obtain the master equation in the standard form of a wave-type equation}
\begin{eqnarray}\label{master_eq}
\frac{\partial^2\Psi}{\partial \tilde{r}_{*}^2}&+&\left[\frac{1}{2\mathcal{S}_0}\frac{\partial^2\mathcal{S}_0}{\partial \tilde{r}_{*}^2}-\frac{3}{4\mathcal{S}_0^2}\left(\frac{\partial\mathcal{S}_0}{\partial \tilde{r}_{*}}\right)^2-\frac{1}{r\mathcal{S}_0}\frac{\partial\mathcal{S}_0}{\partial \tilde{r}^{*}}e^{\mu_0-\nu_0}-\frac{e^{2\mu_0-2\nu_0}}{r^2}(2-r(\mu_0'-\nu_0')) \right. \nonumber \\[3mm]
&&\left. -(l-1)(l+2)\,\frac{e^{2\mu_0}\mathcal{S}_0}{r^2\mathcal{W}_0}\right]\Psi =\frac{\mathcal{S}_0}{\mathcal{P}_0}\partial^2_{t}\Psi \ .
\end{eqnarray}

For the study of QNMs, it is useful to transform the above equation
into the Schr\"odinger form.
For this purpose, we have to introduce a deformed tortoise coordinate $r_*$,
related to the previous $\tilde{r}_*$ coordinate in the following way:
\begin{equation}\label{eq:tortoise}
\frac{d r_*}{d r}=e^{\nu_0-\mu_0}\sqrt{\frac{\mathcal{S}_0}{\mathcal{P}_0}}=\frac{d \tilde{r}_*}{d r}\sqrt{\frac{\mathcal{S}_0}{\mathcal{P}_0}} \ .
\end{equation}
In addition, we parameterize the perturbation variable:
\begin{eqnarray}
\Psi(r,t) = \left(\frac{\mathcal{P}_0}{\mathcal{S}_0}\right)^{1/4}
e^{-i\omega t} Z(r) \ .
\end{eqnarray}
With these definitions and performing simple manipulations,
Eq.~\eqref{master_eq} can be transformed into the Schr\"odinger-like form
\begin{equation}\label{Schrodinger}
\frac{d^2Z}{d r_{*}^2} + (\omega^2 - V_0(r))Z = 0 \ ,
\end{equation}
where $V_0(r)$ is a potential that depends only on the background quantities.

{As we will see below, even though for large $r_{\rm H}$ the function $\mathcal{S}_0$ is everywhere positive. For a certain set of solutions with smaller $r_{\rm H}$ it can have nodes and negative regions outside the horizon. The function $P_0$ on the other hand is always positive.
Therefore, for these solutions the tortoise coordinate $r_{*}$, the potential $V_0$
and the perturbation variable $Z$ become ill-defined. In addition, the perturbation equation \eqref{master_eq} is no longer hyperbolic, which leads to severe problems for
 the axial perturbations of the solutions.}

In order to be able to counter-check our results
and to be more secure about the regions where instabilities appear,
we have implemented two different methods for finding the QNMs
-- performing time evolution of the perturbation Eq.~\eqref{master_eq}
and solving the eigenvalue problem defined by Eq.~\eqref{Schrodinger}.
The relevant boundary conditions one has to impose in both cases
are that the perturbation function has the form of an outgoing wave
at infinity and an ingoing wave at the horizon:
\begin{eqnarray}
\Psi \xrightarrow[r \to \infty]{} e^{-i\omega(t-r_*)} \ , \nonumber \\
\Psi \xrightarrow[r \to r_{\rm H}]{} e^{-i\omega(t+r_*)} \ . \label{eq:BC_in_outgoing}
\end{eqnarray}
Clearly, the QNM frequency $\omega$ is a complex variable,
where the real part $\omega_R$ controls the oscillation frequency,
while the imaginary part $\omega_I$ is connected
to the damping/growth time of the mode.
It can be easily shown that in case of unstable modes with  $\omega_I<0$,
$\omega_R=0$ and the boundary conditions at the two boundaries are just the trivial ones,
i.e.~$\Psi|_{r=\infty}=\Psi|_{r=r_{\rm H}}=0$.

When solving the time-dependent {problem for the stable modes}, one has to impose
the standard radiative boundary conditions derived from
Eq.~\eqref{eq:BC_in_outgoing}:
\begin{eqnarray}
&&{\rm horizon:}\;\;\; \partial_t \Psi - \partial_{r_*} \Psi =0 \ , \nonumber \\
&&{\rm infinity \; :}\;\;\;   \partial_t \Psi + \partial_{r_*} \Psi =0 \ .
\end{eqnarray}

Imposing the boundary conditions Eq.~\eqref{eq:BC_in_outgoing}
when solving the time-independent eigenvalue problem
leads to severe difficulties, because the ingoing (outgoing)
boundary condition is easily contaminated by the undesired
outgoing (ingoing) one when solving the problem numerically.
Thus, special treatment is required.

The method we use for the time-independent problem is based on the previous results
\cite{Blazquez-Salcedo:2016enn,Blazquez-Salcedo:2017txk,Blazquez-Salcedo:2018jnn,Blazquez-Salcedo:2019nwd,Blazquez-Salcedo:2020nhs}.
Let us give a brief summary here.
We compactify and divide the domain of integration into two parts,
one including the horizon, and the other one extending to infinity.
On the first part we parameterize the ingoing wave behavior
and on the second part the outgoing wave behavior,
making use of a power expansion of the solution at each boundary.
We generate solutions of the perturbations for arbitrary values of $\omega$
and determine the QNMs by looking for the solutions
that match smoothly the ingoing perturbation with the outgoing one.
We refer the reader to \cite{Blazquez-Salcedo:2018pxo} for further details.

\section{Results}

In what follows we will work with $\lambda=1$ without loss of generality,
which is equivalent to rescaling the dimensional quantities
with respect to $\lambda$.
Then, after fixing the coupling function \eqref{eq:coupling_function}
and setting the scalar field potential to zero,
the black holes can be parametrized by the value
of the horizon radius $r_{\rm H}$.
The domain of existence and the properties of the solutions
were discussed in detail in \cite{Doneva:2017bvd,Blazquez-Salcedo:2018jnn},
and that is why we will comment on them only briefly here.

The particular form of $f(\varphi)$ given by Eq.~\eqref{eq:coupling_function}
(quadratic leading-order coupling)
allows for spontaneous scalarization.
This means that the Schwarzschild black hole
is always a solution of the field equations,
but when the curvature of the spacetime reaches a certain threshold value,
the Schwarzschild solution loses stability
under spherically symmetric perturbations,
and a new branch of scalarized black holes bifurcates from it.
In fact, more than one branch exist,
and the additional branches bifurcate at values of $r_{\rm H}$
where new unstable modes of the Schwarzschild solution appear.
The additional branches can be labeled by the number of nodes
of the scalar field.
All the solutions with nodes are unstable
against radial perturbations \cite{Blazquez-Salcedo:2018jnn}.
{The fundamental branch, though, that is characterized by black hole solutions possessing no nodes of the scalar field,} was shown to be stable under
radial perturbations for a large range of values of the
horizon radius.

\begin{figure}
	\centering
	\includegraphics[width=0.34\linewidth,angle=-90]{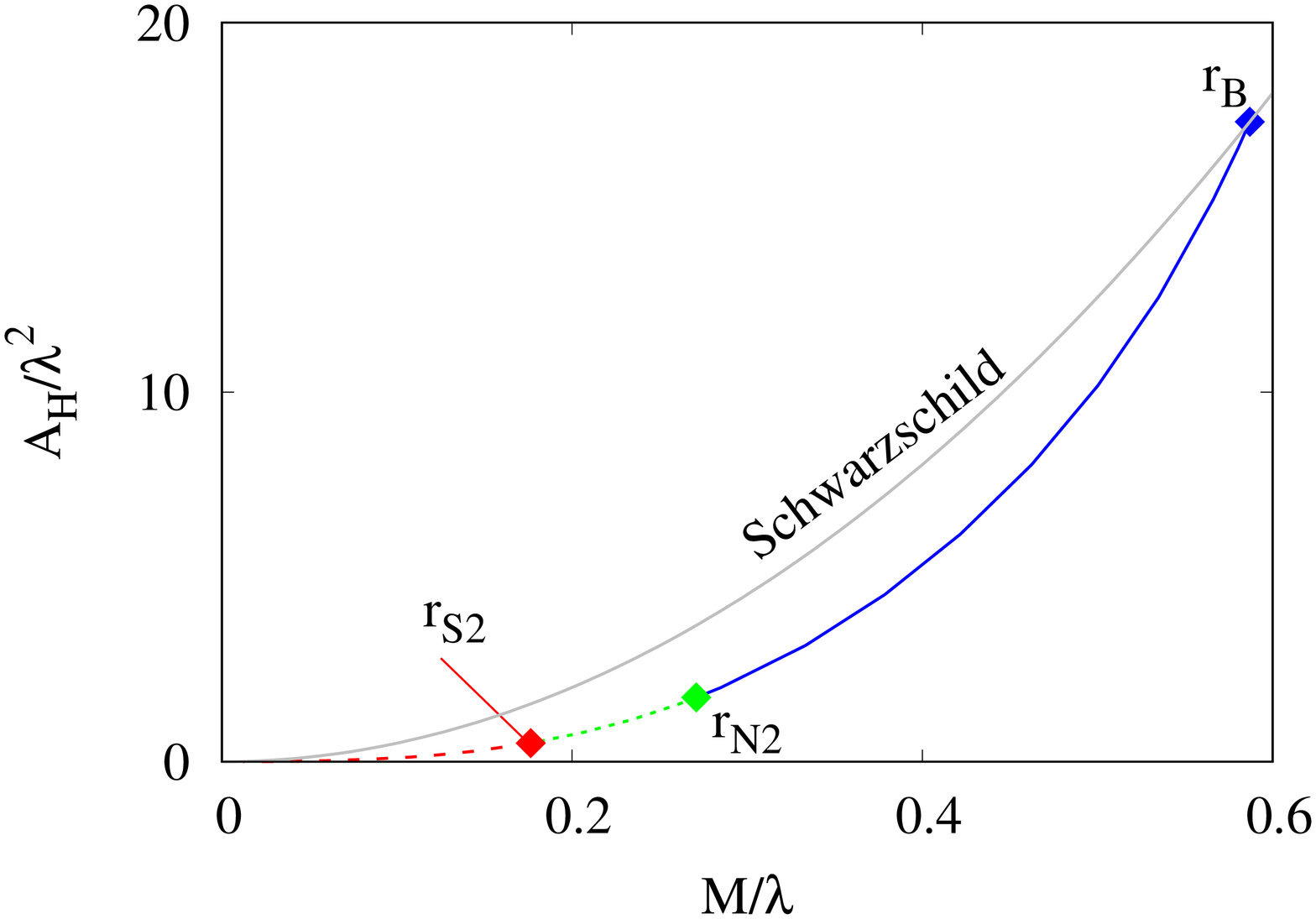}
	\includegraphics[width=0.34\linewidth,angle=-90]{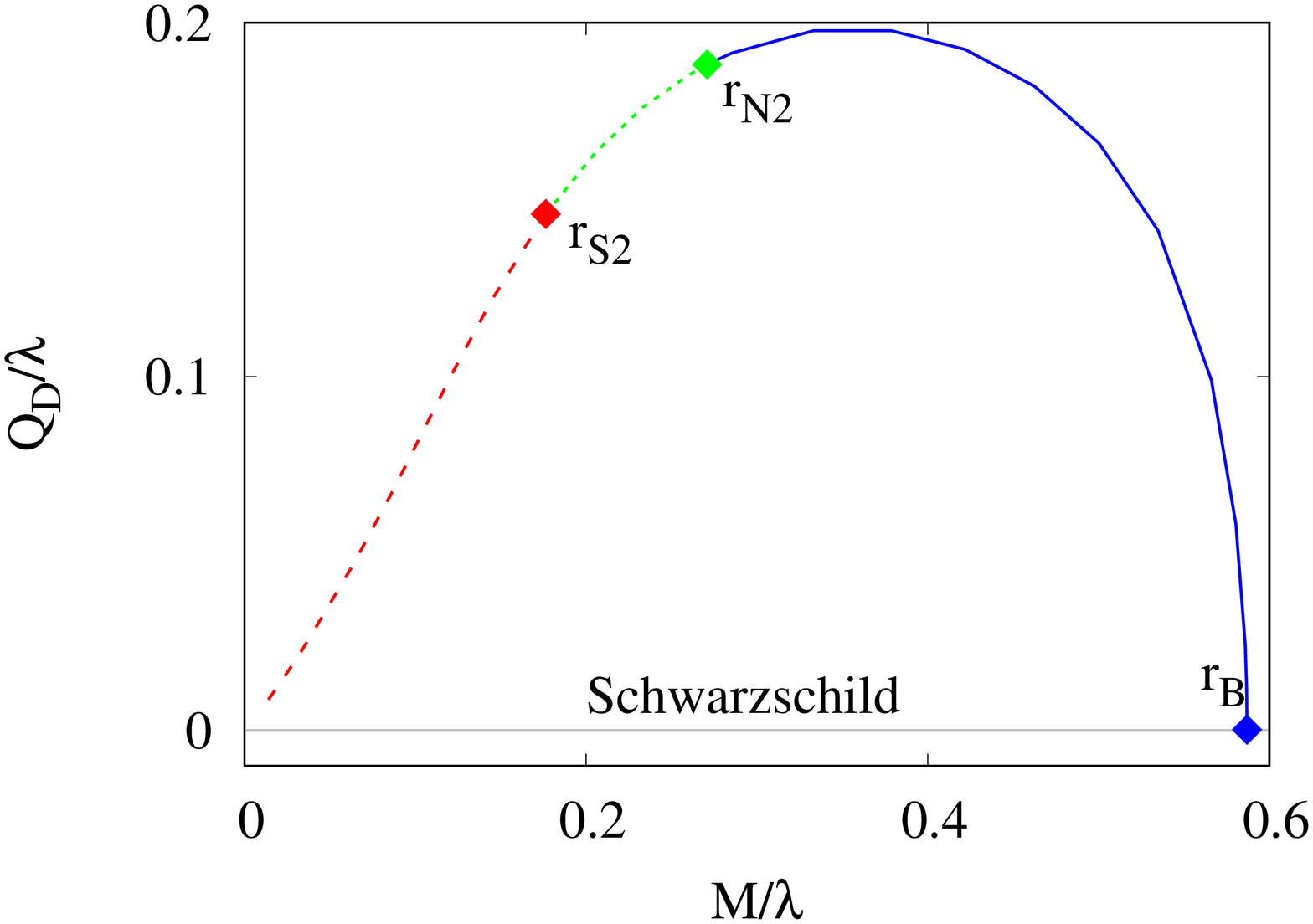}
	\caption{(left) Scaled horizon area $A_{\rm H}/\lambda^2$
vs. scaled total mass $M/\lambda$.
(right) Scaled scalar charge $Q_{\rm D}/\lambda$ vs. scaled total mass.
The bifurcation point is denoted by $r_{\rm B}$. In the range
$r_{\rm B} > r_{\rm N2}$ (solid blue) the axial potential is strictly positive.
In the range $r_{\rm B} > r_{\rm N2}$ (dotted green) the potential is no
longer strictly positive, but the solutions are still stable with respect
to axial perturbations.
In the range $r_{\rm S2} \ge r_{\rm H} > 0$ (dashed red) hyperbolicity is lost.
For comparison also the Schwarzschild solution is shown (solid grey).}
	\label{fig:static}
\end{figure}

We exhibit the fundamental branch together with the Schwarzschild branch
in Fig.~\ref{fig:static},
where we show the horizon area $A_{\rm H}$ (left)
and the scalar charge $Q_{\rm D}$ (right)
as a function of the mass $M$, where all quantities are scaled
with respect to the coupling parameter $\lambda$.
The fundamental branch bifurcates from the Schwarzschild branch
at $r_{\rm B}=1.173944$ and continues to exist for all
$r_{\rm H}<r_{\rm B}$.
Loss of hyperbolicity and stability of the solutions are
discussed in the following subsections.

\subsection{Loss of hyperbolicity of the small scalarized black holes}

{In \cite{Blazquez-Salcedo:2018jnn} we showed that the solutions with $r_{\rm H} > r_{\rm S1}=0.191605$ are stable with respect to radial perturbations, even though the radial potential was not strictly positive for black holes with $r_{\rm S1} < r_{\rm H} < r_{\rm N1}=0.406$.} For solutions with $r_{\rm H} < r_{\rm S1}$, {the tortoise coordinate becomes ill-defined, indicating the loss of hyperbolicity of the radial perturbation equation.}

\begin{figure}
	\centering
	\includegraphics[width=0.38\linewidth,angle=-90]{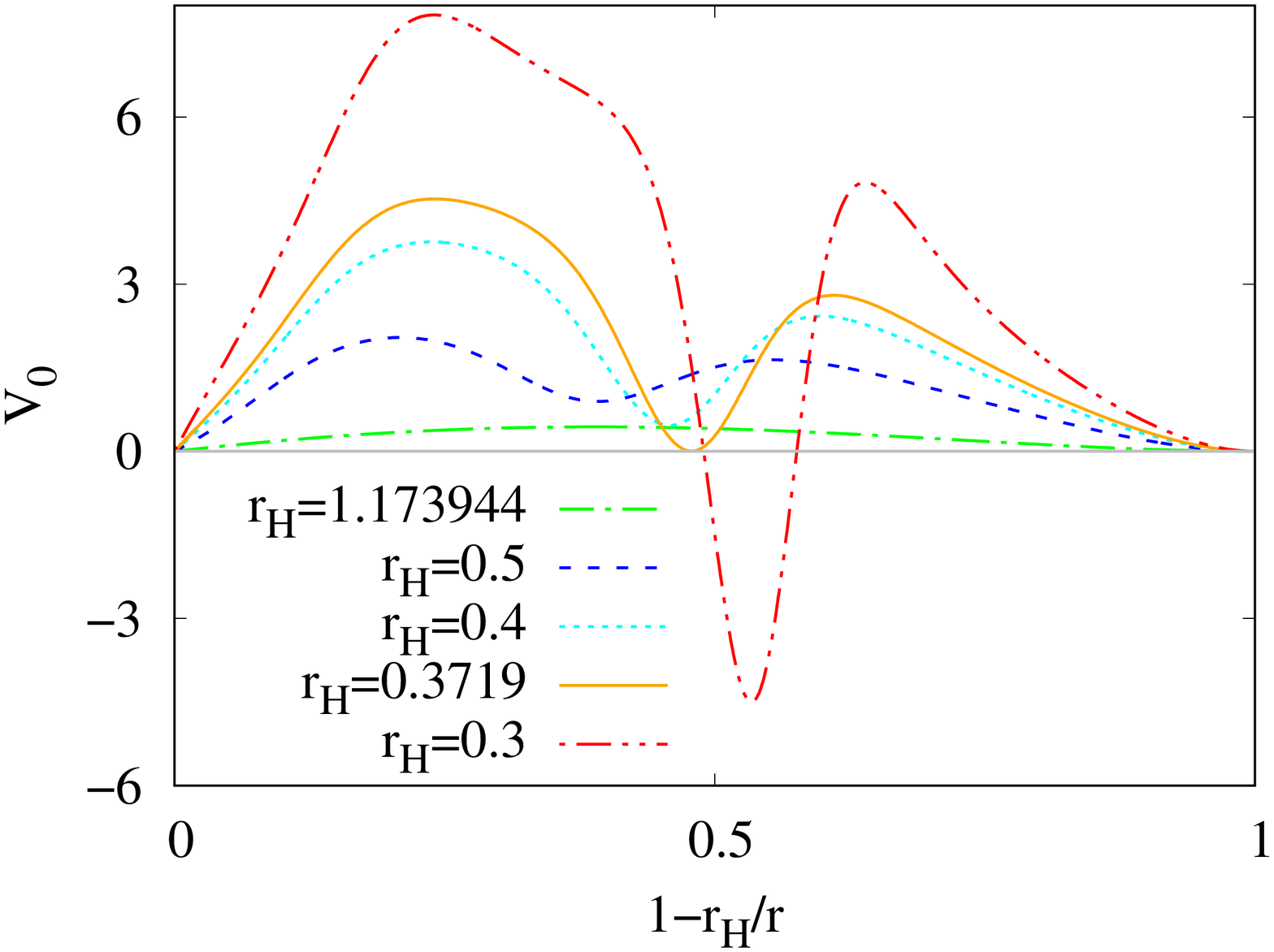}
	\includegraphics[width=0.38\linewidth,angle=-90]{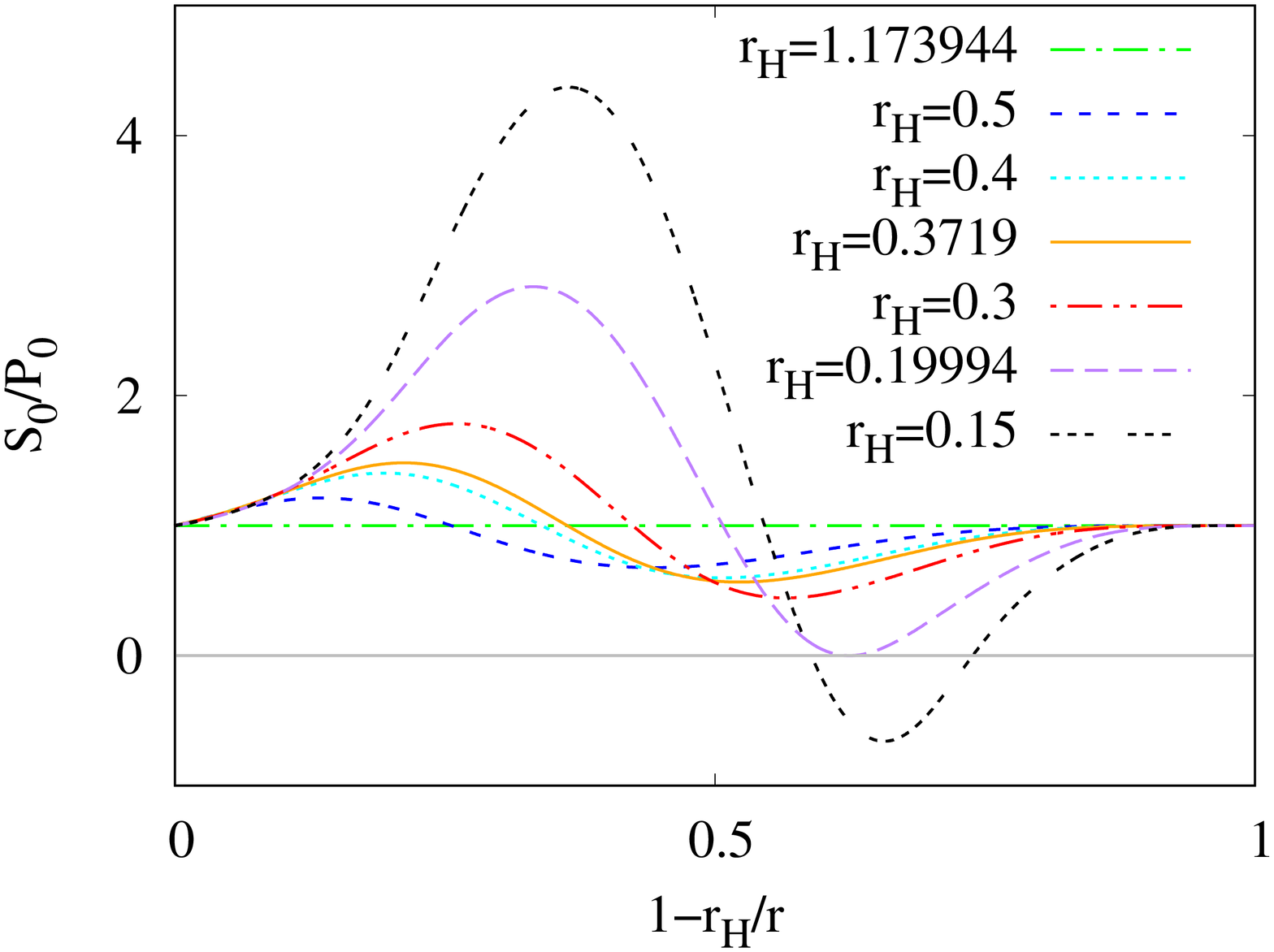}
	\caption{(left) Effective axial potential $V_0$ vs.
compactified coordinate $1-r_{\rm H}/r$ for $l=2$ QNMs
and several values of the horizon radius $r_{\rm H}$.
(right) $\mathcal{S}_0/\mathcal{P}_0$ vs. compactified coordinate
$1-r_{\rm H}/r$.}
	\label{fig:plotpotential1_and_W}
\end{figure}

Let us now focus on the axial perturbations.
It turns out that the situation is similar
to what happens with the radial perturbations.
However, the range of parameters is slightly changed.
As one can see in the left panel of Fig.~\ref{fig:plotpotential1_and_W},
{where the radial profile of the axial potential $V_0$ is shown for $l=2$ and several representative black hole solutions},
$V_0$ is strictly positive for $r_{\rm B} > r_{\rm H} > r_{\rm N2}=0.3718$,
and thus these black holes are in any case stable under axial perturbations.
We have marked these black holes in Fig.~\ref{fig:static} by a solid blue line.
The potential is no longer strictly positive for the solution with
$r_{\rm H}=r_{\rm N2}$.
We have marked this limiting solution by a green dot
in Fig.~\ref{fig:static}.
In fact, the potential is negative in some intermediate region
outside the horizon for solutions with
$r_{\rm N2} > r_{\rm H} > r_{\rm S2}=0.19994$.
We have marked these solutions by a dotted green line
in Fig.~\ref{fig:static}.

For even lower values of $r_{\rm H}$ the situation is more complex
because the potential and the tortoise coordinate become ill-defined.
Motivated by Eq.~\eqref{eq:tortoise},
we have plotted the function $\mathcal{S}_0/\mathcal{P}_0$
in the right panel of Fig.~\ref{fig:plotpotential1_and_W}.
Similar to the radial perturbations,
this function possesses a negative minimum below
$r_{\rm H}=r_{\rm S2}=0.19994$.
It is interesting to note that the value of the horizon radius $r_{\rm S2}$,
where this happens for the axial perturbations,
is different from the corresponding one for the radial perturbations,
$r_{\rm S1}$, with $r_{\rm S2} > r_{\rm S1}$.
Clearly, for $r_{\rm H} < r_{\rm S2}$ the tortoise coordinate
of the axial perturbations becomes ill-defined,
and this does not depend on the angular $l$ number of the perturbations
because the function $\mathcal{S}_0/\mathcal{P}_0$
depends only on the background quantities.

{Since this function controls the sign of the time derivative
of the axial perturbation wave equation (\ref{master_eq}), it is no longer hyperbolic
for all the solutions with $r_{\rm H} < r_{\rm S2}$.
This is then a slightly larger set of solutions
than those leading to a non-hyperbolic radial perturbation equation. }
We have marked the $r_{\rm S2}$ solution in Fig.~\ref{fig:static}
by a red dot, and
{all other solutions that lead to a non-hyperbolic axial perturbation equation}
are marked by a dashed red line.
{Note that this line includes all the solutions
that are not hyperbolic under radial perturbations.}
{As seen in Fig.~\ref{fig:static},
these solutions with non-hyperbolic perturbation equations
correspond to scalarized black holes having both small masses and small horizon areas.}

In the following, we will focus on
solutions in the range $r_{\rm H} > r_{\rm S2}$,
which are expected to contain the set of stable physically relevant solutions.

\subsection{Stability and the S-deformation method}

Before calculating the spectrum of QNMs of these scalarized black holes,
let us first discuss the subset of solutions
that possess a regular but not strictly positive axial potential,
i.e. the solutions in the range $r_{\rm S2} < r_{\rm H} < r_{\rm N2}$,
represented by the dotted green curve in Fig.~\ref{fig:static}.
In the following, we will show that these solutions are nonetheless
stable under axial perturbations.

For this purpose, we make use of the S-deformation method,
introduced in the context of black holes in
\cite{Kimura:2017uor,Kimura:2018whv}
and used recently for charge-induced spontaneously scalarized
black hole solutions in \cite{Blazquez-Salcedo:2020nhs}.
In fact, it turns out that in order to show the stability of these solutions
under axial perturbations, it is sufficient to show
that a deformation function $S(r)$ exists such that the potential $V_0$
can be deformed to
\begin{eqnarray}
\hat{V}_0=V_0 + \frac{dS}{dr^*} - S^2 \geq 0 \ .
\end{eqnarray}
In particular, finding a solution of this equation with $\hat{V}_0=0$
allows to conclude that the original potential,
although not strictly positive, does not contain unstable modes.
For details on the method we refer the reader to
\cite{Kimura:2017uor,Kimura:2018whv,Blazquez-Salcedo:2020nhs}.

\begin{figure}
	\centering
	\includegraphics[width=0.38\linewidth,angle=-90]{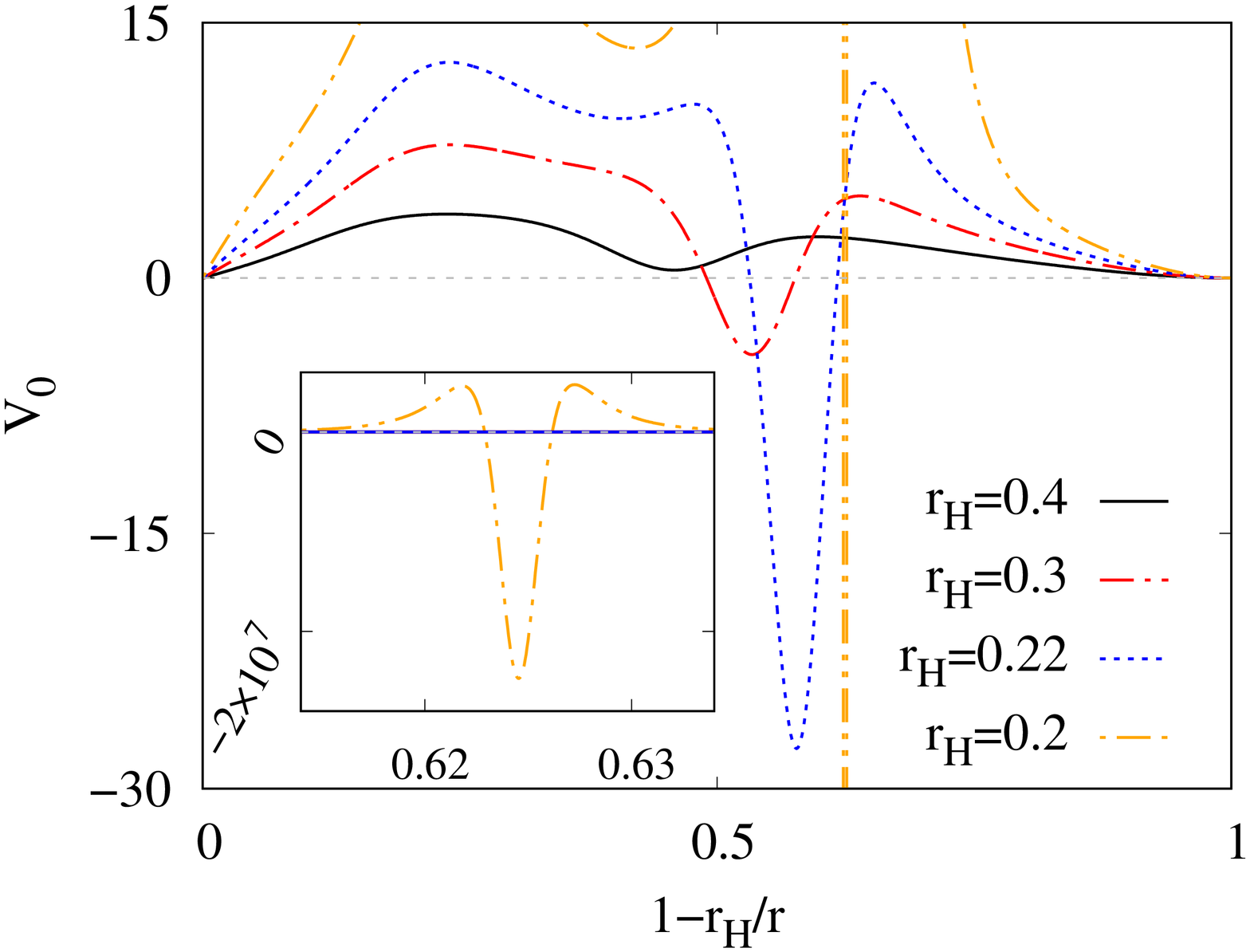}
	\includegraphics[width=0.38\linewidth,angle=-90]{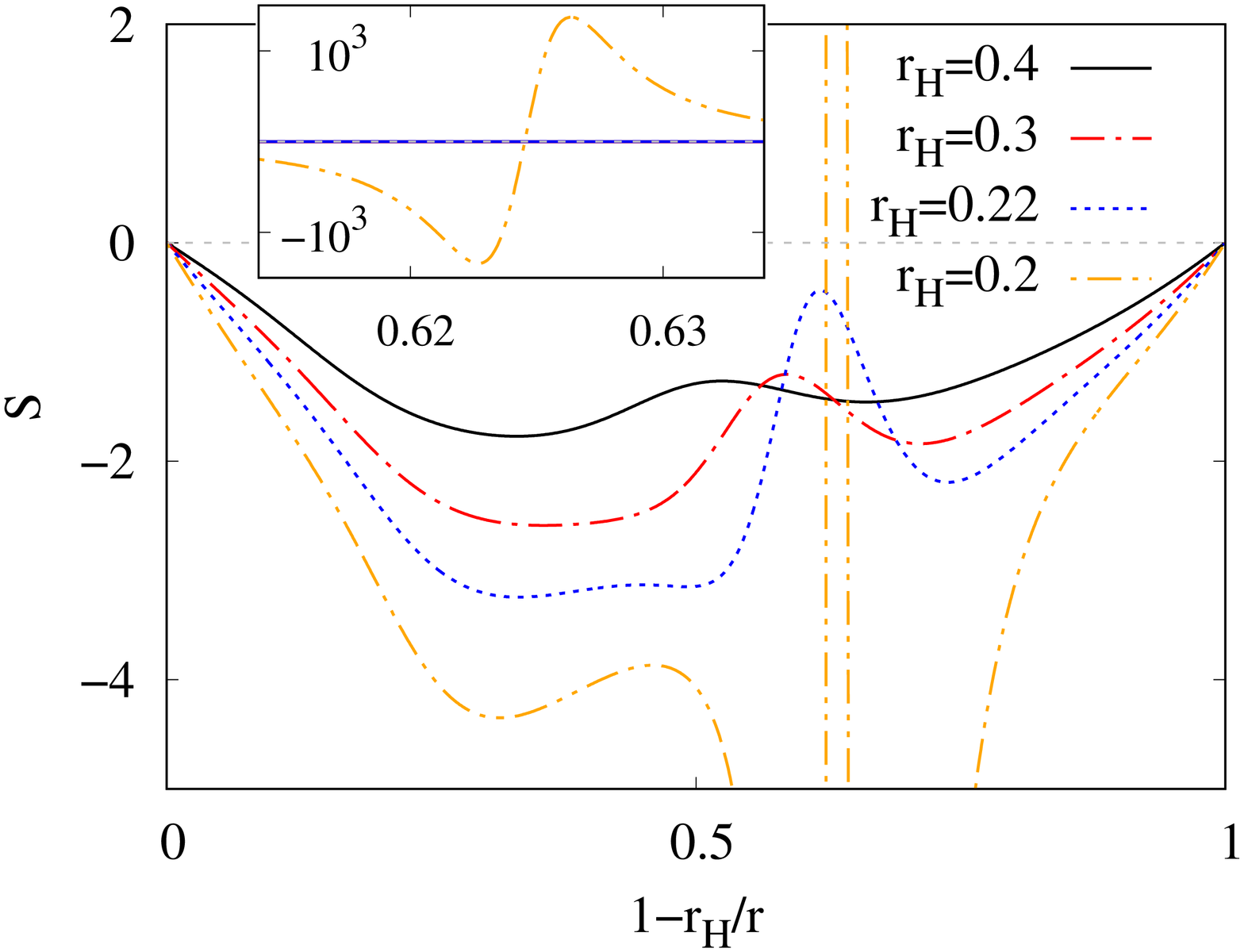}
	\caption{(left) Effective axial potential $V_0$ vs.
compactified coordinate $1-r_{\rm H}/r$ for $l=2$ QNMs
and several values of the horizon radius $r_{\rm H}$
in the range $r_{\rm N2} > r_{\rm H} > r_{\rm S2}$.
(right) Analogous figure for the corresponding S-deformation function
$S$.}
	\label{fig:Sdef}
\end{figure}

Hence, we need to solve the equation
\begin{eqnarray}
\frac{dS}{dx}=\frac{r^2}{r_{\rm H}}\frac{dr^*}{dr}(S^2-V_0) \ ,
\label{eq_S_2}
\end{eqnarray}
where $x=1-r_{\rm H}/r$. Since the potential is zero at $r=r_{\rm H}$ ($x=0$)
and $r=\infty$ ($x=1$), we require $S(x=0)=S(x=1)=0$. 
With these boundary conditions, it is possible
to integrate Eq.~(\ref{eq_S_2}) numerically for all solutions with
not strictly positive potentials $V_0$
in the range $r_{\rm H}>r_{\rm S2}$.
We show some examples in Fig.~\ref{fig:Sdef}.
In the left panel we exhibit several examples of potentials,
and in the right panel we show the corresponding S-deformation function $S$,
that is obtained after integrating Eq.~(\ref{eq_S_2}).
Non-surprisingly, the closer the solution approaches $r_{\rm H} = r_{\rm S2}$,
the steeper the S-function becomes.
However, the S-function remains everywhere smooth
as long as $r_{\rm H} > r_{\rm S2}$. 

From these results we conclude that all sufficiently large
scalarized black holes on the fundamental branch,
i.e. those in the range $r_{\rm H} > r_{\rm S2}$,
are stable under axial perturbations.
These solutions are represented by the dotted green line
and by the solid blue line in Fig.~\ref{fig:static}.
Recall that all these solutions are also stable under radial perturbations,
as shown in \cite{Blazquez-Salcedo:2018jnn}.

\subsection{Quasinormal modes}

Having established that all the fundamental scalarized black holes
with {hyperbolic axial perturbation equation} are stable
(with respect to radial and axial perturbations),
we will now present the spectrum of QNMs for these solutions,
focussing on the fundamental $l=2$ mode.
As described in the previous section,
we obtain the QNMs by employing two independent methods --
the time evolution of the perturbation equation \eqref{master_eq},
and the time-independent eigenvalue problem defined by Eq.~\eqref{Schrodinger}.
The first one is a more direct approach,
but in practice only the first QNM frequency
can be extracted with good accuracy.
The method can still serve as a verification of the eigenvalue results,
especially if there are instabilities.

\begin{figure}
	\centering
	\includegraphics[width=0.46\linewidth]{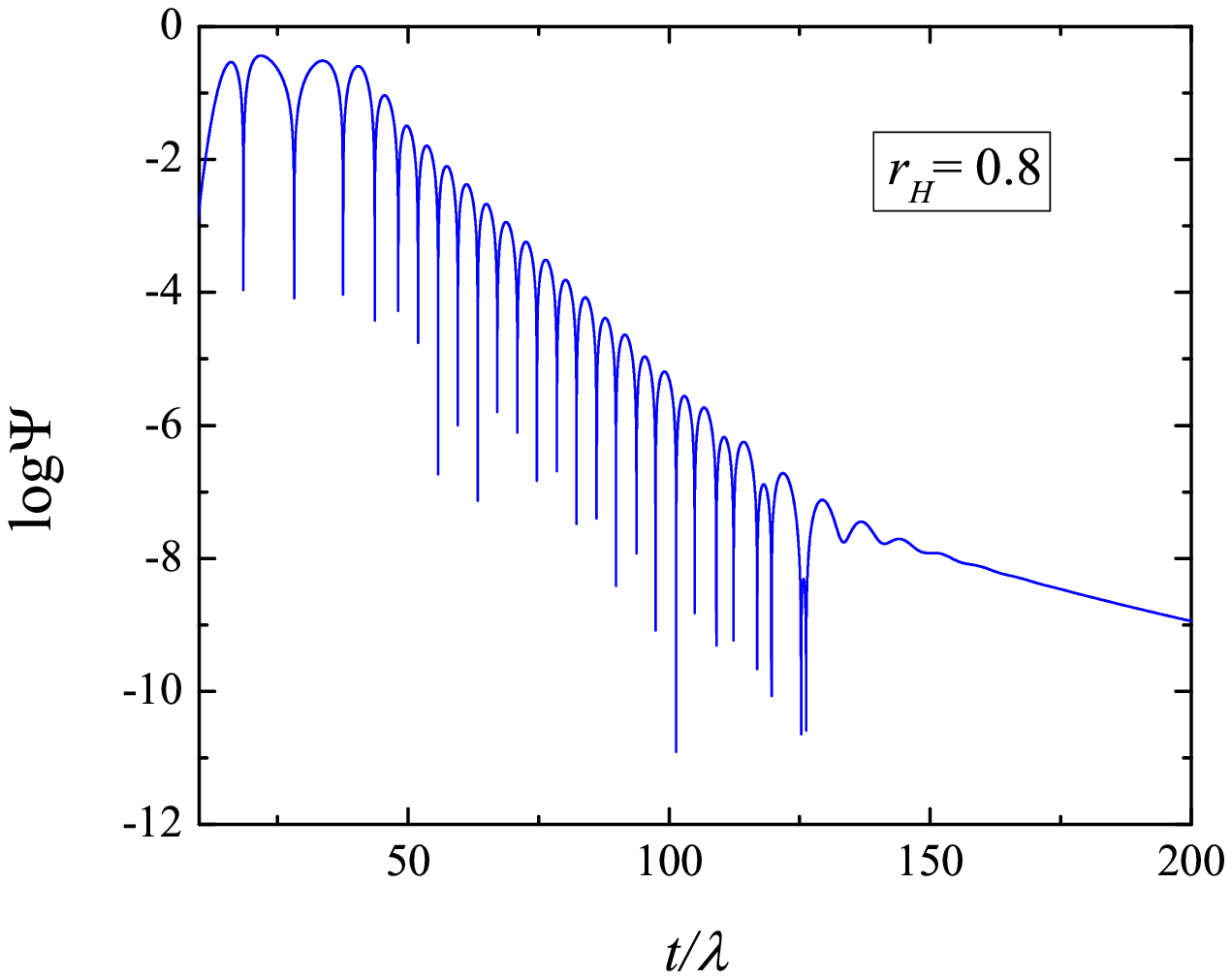}
	\includegraphics[width=0.46\linewidth]{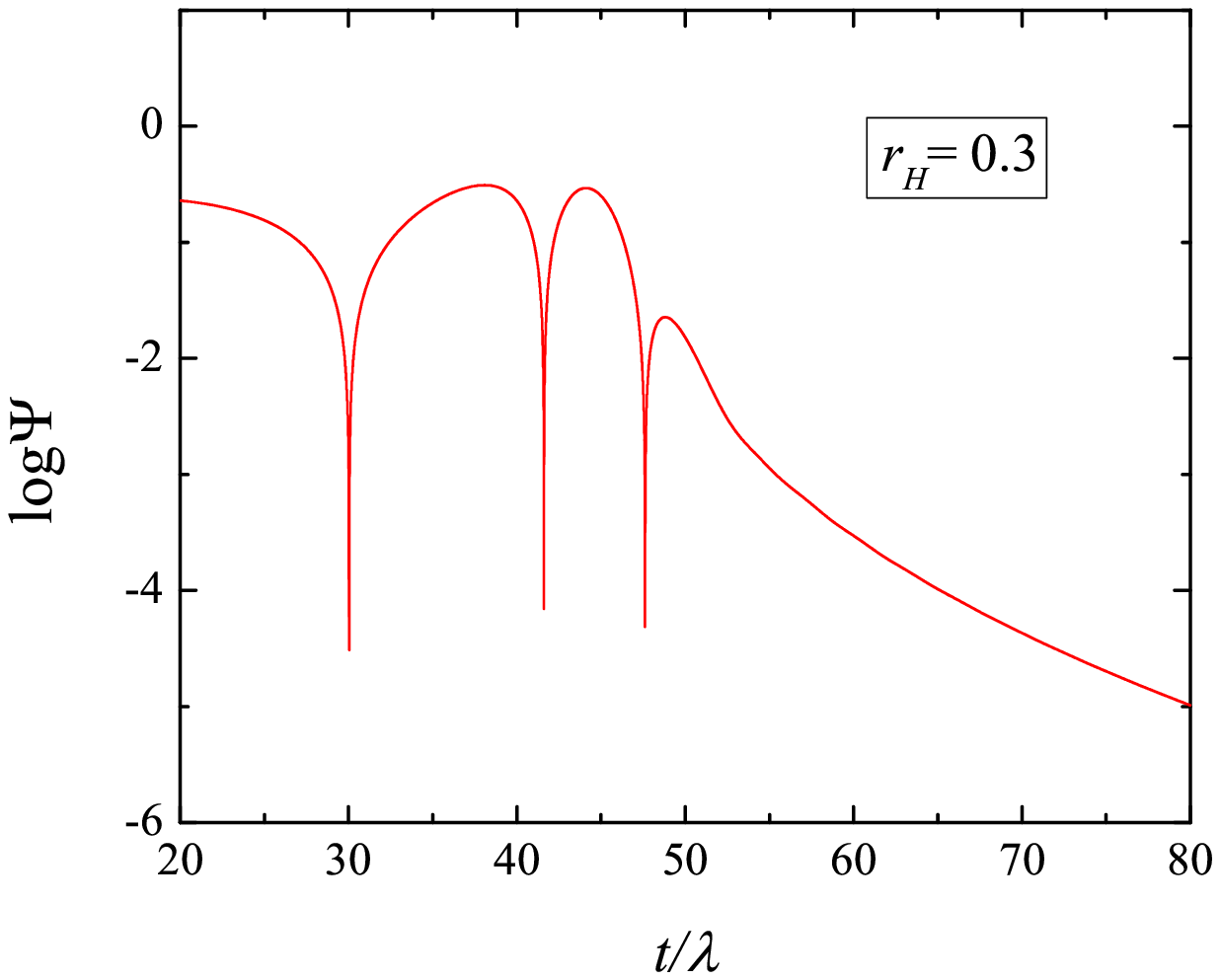}
	\caption{(left) Time evolution of the perturbation $\Psi$
for logarithmic scale observed at coordinate distance $r=10$
(in geometrical units) for $r_{\rm H}=0.8$, where the axial potential $V_0$
is strictly positive.
(right) Time evolution for $r_{\rm H}=0.3$, where the potential
has a negative minimum.}
	\label{fig:time_evolution}
\end{figure}

The time evolution of the perturbation $\Psi$,
starting with a Gaussian pulse as an initial data,
is presented in Fig.~\ref{fig:time_evolution}
for two black hole solutions.
The first choice in the left panel with $r_{\rm H}=0.8$
corresponds to a solution possessing a potential $V_0(r)$
that is everywhere strictly positive.
For the black hole with $r_{\rm H}=0.3$ the potential has a negative minimum.
The signal is observed at a coordinate distance of $r=10$
in dimensionless units.
As one can see, the perturbations are stable in both cases,
leading to QNM oscillations followed by a power-law tail.
The comparison of the extracted QNM frequencies with those
obtained with the time-independent eigenvalue code
(for solutions with a hyperbolic equation)
shows very good agreement, with a maximum discrepancy of roughly $1.5\%$.

The time evolution shows the development of an instability for black holes with $r_{\rm H}<0.202$ that is just 1\% larger than the value of $r_{\rm S2}$ where the perturbation equation loses its hyperbolic character. This discrepancy is within the numerical error and is due to the formation of a very deep and sharp minimum for black holes with $r_{\rm H}$ close to  $r_{\rm S2}$ that is very difficult to be treated numerically.

\begin{figure}
	\centering
	\includegraphics[width=0.34\linewidth,angle=-90]{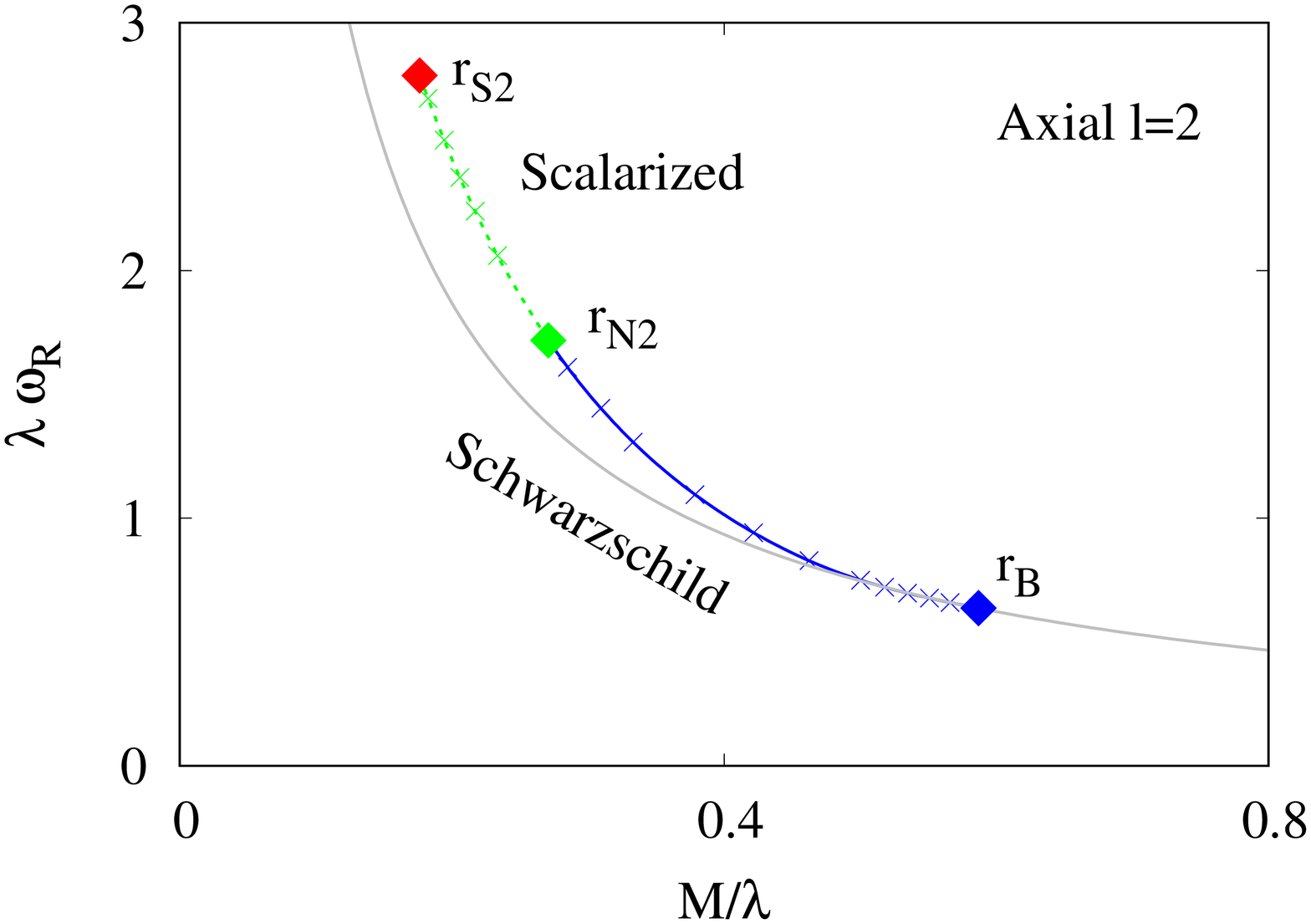}
	\includegraphics[width=0.34\linewidth,angle=-90]{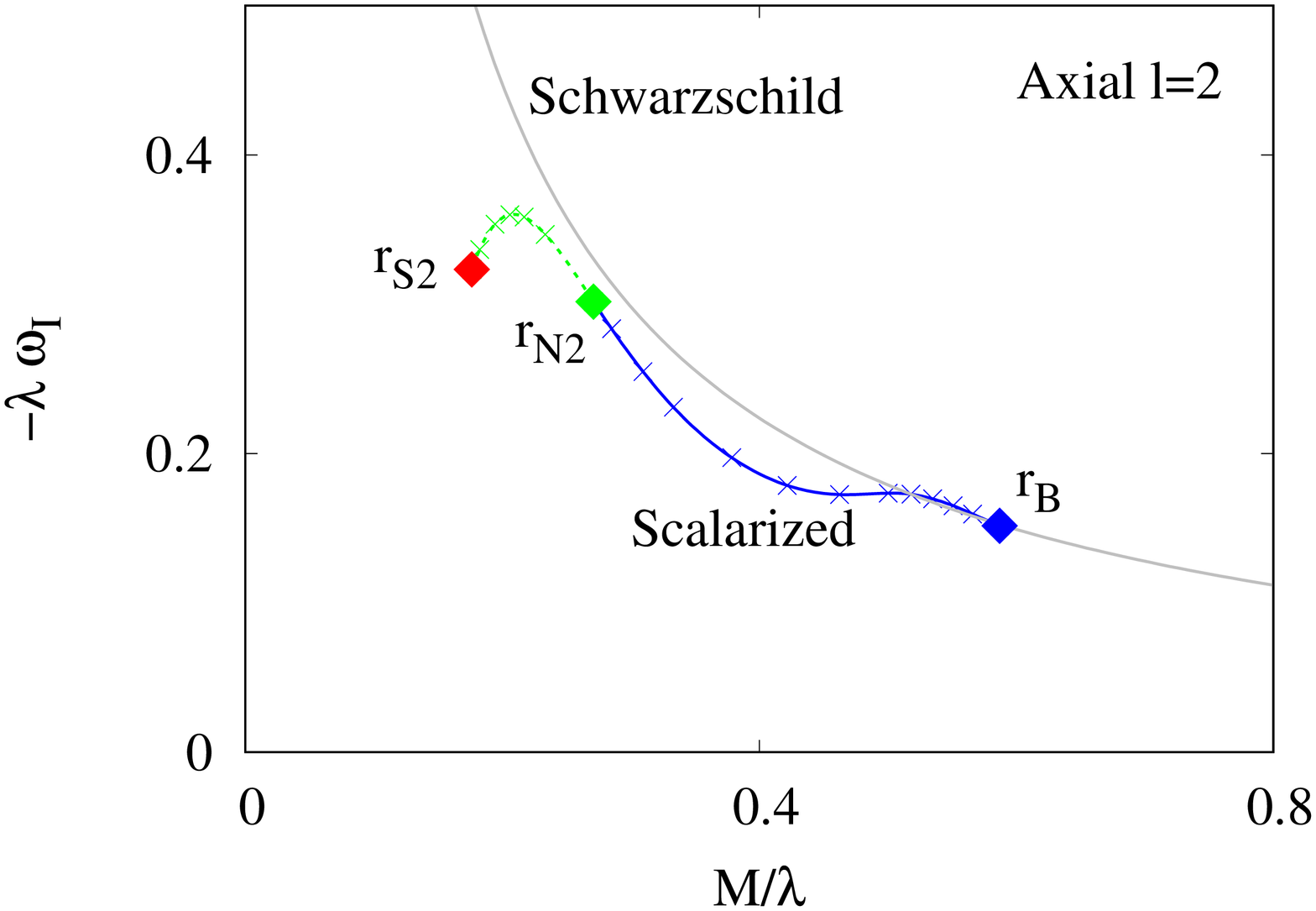}
	\caption{
Scaled axial $l=2$ QNM eigenvalue $\lambda \omega$
vs. scaled total mass $M/\lambda$: real part/frequency $\omega_R$
(left) and imaginary part/inverse damping time $\omega_I$ (right).
The bifurcation point from Schwarzschild is denoted by $r_{\rm B}$.
In the range $r_{\rm B} > r_{\rm N2}$ (solid blue)
the axial potential is strictly positive.
In the range $r_{\rm B} > r_{\rm N2}$ (dotted green) the potential is no
longer strictly positive, but the solutions are still stable with respect
to axial perturbations.
Beyond $r_{\rm S2}$ (red) the hyperbolicity is lost.
For comparison also the Schwarzschild QNM frequencies are shown (solid grey).}
	\label{fig:plotmomegawlambda}
\end{figure}

\begin{figure}
	\centering
	\includegraphics[width=0.34\linewidth,angle=-90]{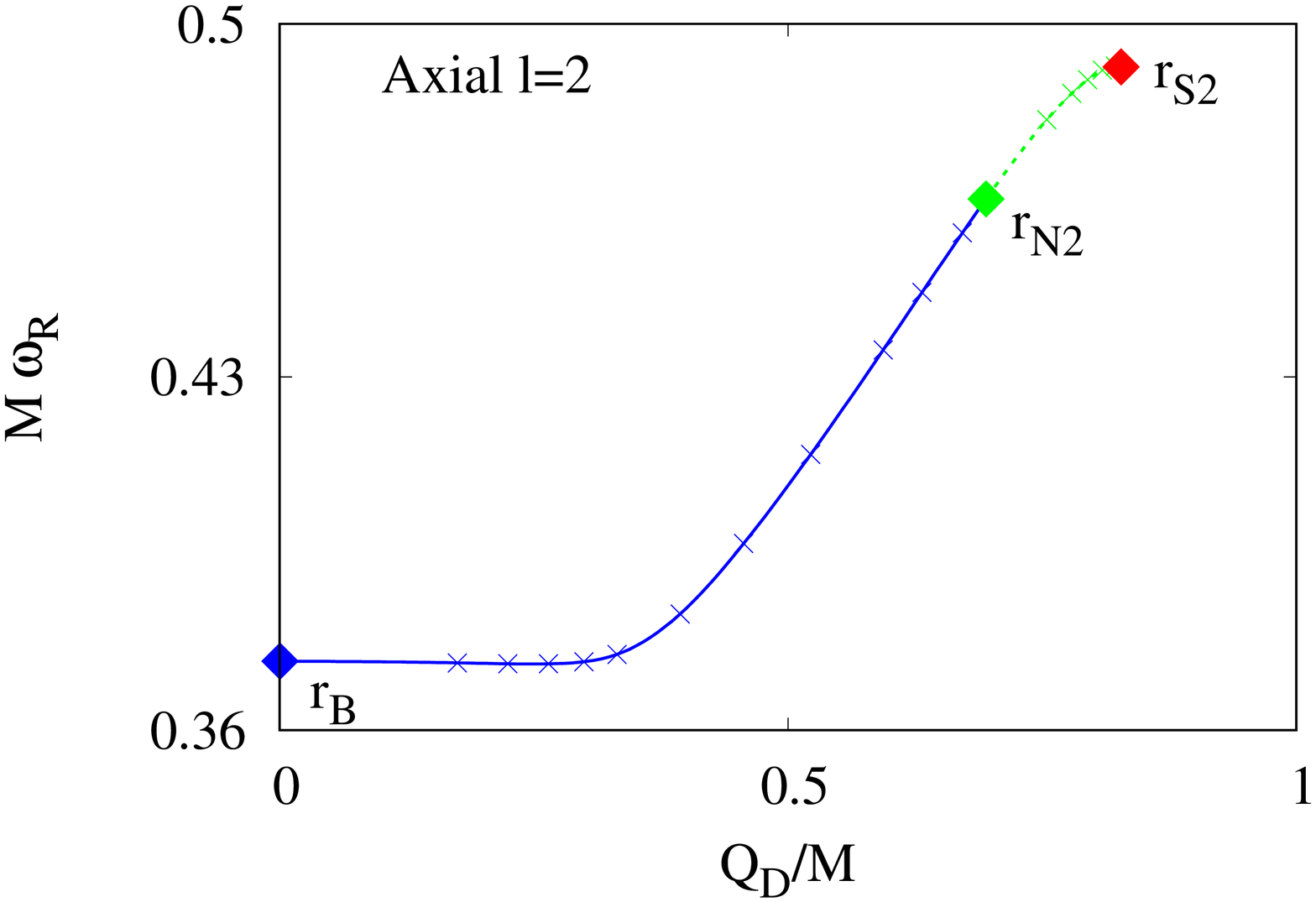}
	\includegraphics[width=0.34\linewidth,angle=-90]{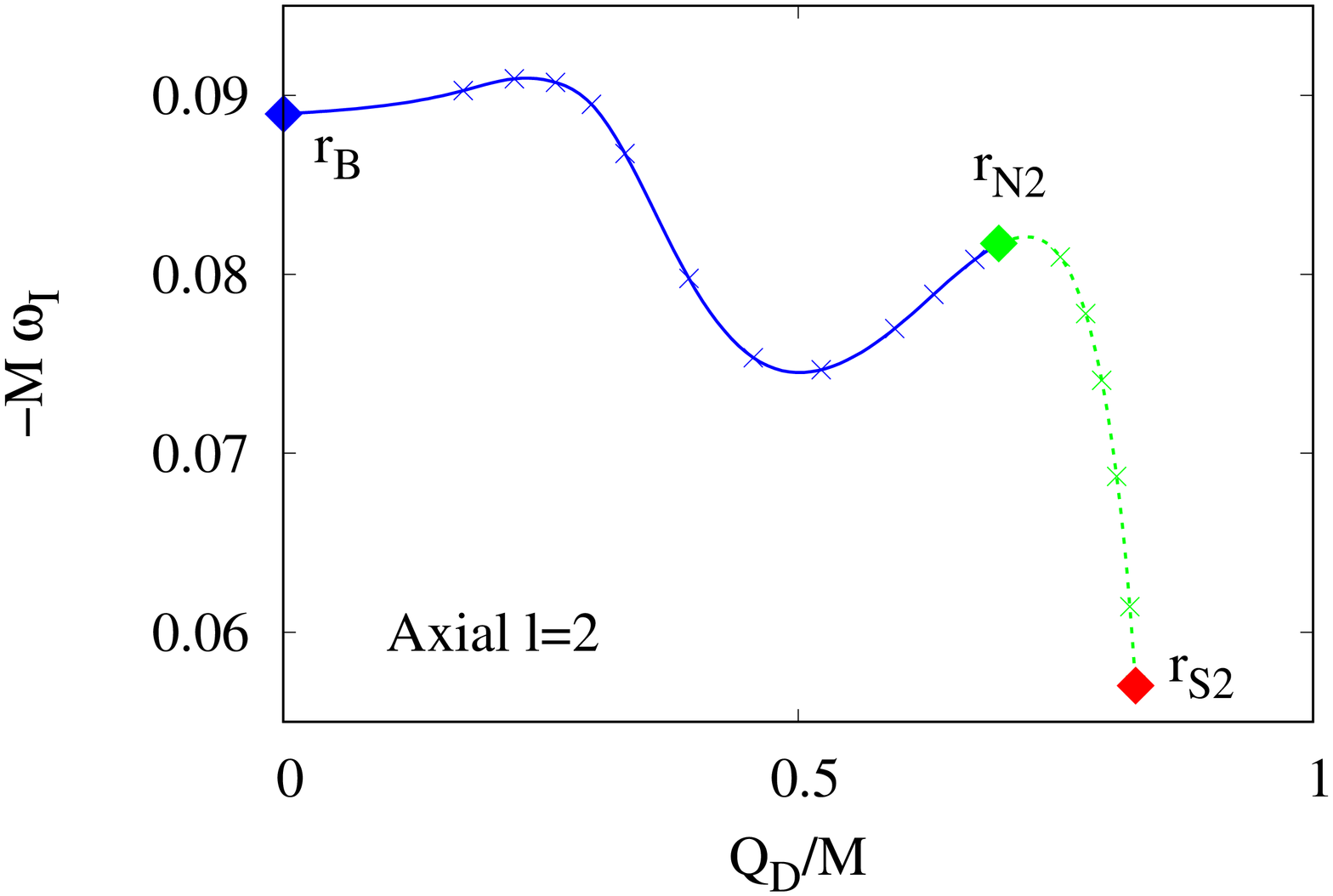}
	\caption{
Scaled axial $l=2$ QNM eigenvalue $M \omega$
vs. scaled scalar charge $Q_{\rm D}/M$: real part/frequency $\omega_R$
(left) and imaginary part/inverse damping time $\omega_I$ (right).
The bifurcation point from Schwarzschild is denoted by $r_{\rm B}$.
In the range $r_{\rm B} > r_{\rm N2}$ (solid blue)
the axial potential is strictly positive.
In the range $r_{\rm B} > r_{\rm N2}$ (dotted green) the potential is no
longer strictly positive, but the solutions are still stable with respect
to axial perturbations.
Beyond $r_{\rm S2}$ (red) the hyperbolicity is lost.}
	\label{fig:plotQdomega}
\end{figure}

Let us now consider the results from the time-independent method.
Starting at $r_{\rm H} = r_{\rm B}$
and tracking the fundamental axial $l=2$ mode
of the Schwarzschild solution ($M\omega=0.3736-i0.0895$),
we generate the modes shown in Fig.~\ref{fig:plotmomegawlambda}.
The figure shows the scaled eigenvalue $\lambda \omega$
versus the scaled total mass $M/\lambda$.
On the left the real part of the mode is shown,
representing the frequency $\omega_R$,
while on the right the imaginary part is shown,
representing the inverse damping time $\omega_I$.
Also the critical horizon radii $r_{\rm B}$,
$r_{\rm N2}$ and $r_{\rm S2}$ are indicated.
The modes are shown with the color coding
$r_{\rm B} > r_{\rm H} > r_{\rm N2}$ (solid blue) and
$r_{\rm N2} > r_{\rm H} > r_{\rm S2}$ (dotted green),
and for comparison also the Schwarzschild modes are shown (solid grey).
We observe that the scaled frequency of the scalarized black holes
of a given mass is always slightly larger than the corresponding frequency
of the Schwarzschild black holes.
The imaginary part of the scalarized black holes
is always slightly smaller than their Schwarzschild counterpart,
meaning that the damping time is slightly larger for the scalarized solutions.

In Fig.~\ref{fig:plotQdomega} we show the fundamental axial $l=2$ mode
versus the scalar charge $Q_D$ of the black holes,
where all quantities are scaled with respect to the black hole mass $M$.
Here we see that the frequency remains almost constant
for small and intermediate values of the scalar charge,
and it only starts to grow significantly
when the scalar charge is sufficiently large.
In contrast, the imaginary part has a more sinusoidal dependence on
the scalar charge, with an overall tendency to decrease
for sufficiently large values of the scalar charge.

We have checked that there are no unstable modes for solutions
in the range $r_{\rm S2} < r_{\rm H} < r_{\rm N2}$.
This is compatible with the existence of the S-deformation function
as discussed in the previous section.
Finally, let us note that by naively integrating
the time-independent equations, it seems possible to obtain unstable modes
for the non-hyperbolic equations in the range $r_{\rm H} < r_{\rm S2}$,
forming a tower of purely imaginary modes with $\omega_I>0$.
These modes seem to exist in the full range $r_{\rm H} < r_{\rm S2}$,
and diverge when $r_{\rm H} \to r_{\rm S2}$.
This is similar to the results obtained for the radial perturbations
\cite{Blazquez-Salcedo:2018jnn}.
Let us note that such an instability is also observed when using
naively the time evolution method in the range $r_{\rm H} < r_{\rm S2}$.

\section{Conclusions}

Motivated by current and future gravitational wave
observations from black hole mergers and the need to provide
predictions from promising alternative theories of gravity,
we have studied axial perturbations of spontaneously scalarized
black holes in EGB theory.
We have focussed on black holes obtained with a coupling function
that allows for a fundamental branch of scalarized black holes that are
stable under radial perturbations in a large part of their domain
of existence \cite{Blazquez-Salcedo:2018jnn}.
We have then analyzed the QNMs by two different methods,
by the time evolution method and by the direct integration
of the time-independent Schr\"odinger-like master equation.

When analyzing the axial potential of the master equation for
the QNMs, we have noted that the axial potential is
strictly positive only in the range $r_{\rm N2} < r_{\rm H} < r_{\rm B}$,
whereas strict positivity is lost for $r_{\rm S2} < r_{\rm H} < r_{\rm N2}$.
By making use of the S-deformation method,
we have then demonstrated that the scalarized solutions
are stable with respect to axial perturbations in the full range
$r_{\rm S2} < r_{\rm H} < r_{\rm B}$,
a fact that was confirmed independently by both methods employed for the
calculation of the axial QNMs.

Similar to the case of the radial perturbation equations, which
lose hyperbolicity {at a horizon radius }$r_{\rm S1}$, the
axial perturbation equations lose hyperbolicity at $r_{\rm S2}$,
where $r_{\rm S2}$ is slightly larger than $r_{\rm S1}$.
{Therefore, all scalarized solutions} that are stable with respect
to axial perturbations are also stable with respect to
radial perturbations (but not vice versa).

The eigenvalues of the fundamental $l=2$
axial QNMs of the spontaneously scalarized black holes
are not too different from those of the
corresponding Schwarzschild solutions with the same mass.
In fact, the frequency and the damping time are always slightly larger
for the scalarized black holes as compared to the Schwarzschild black holes.

In order to learn about the full mode stability of these
scalarized black holes, also the (non-radial) polar QNMs
need to be investigated. This represents work in progress.
Here the scalar perturbations no longer decouple,
resulting in a much richer spectrum,
similar to the case of dilatonic EGB black holes
\cite{Pani:2009wy,Blazquez-Salcedo:2016enn,Blazquez-Salcedo:2017txk,Blazquez-Salcedo:2018pxo,Konoplya:2019hml,Zinhailo:2019rwd}.

\section*{Acknowledgements}
JLBS, SK, JK and PN gratefully acknowledge support by the DFG funded
Research Training Group 1620 ``Models of Gravity''.
JLBS would like to acknowledge support from the DFG project BL 1553.
DD acknowledges financial support via an Emmy Noether Research Group funded by the German Research Foundation (DFG) under grant
no. DO 1771/1-1. DD is indebted to the Baden-W\"urttemberg Stiftung for the financial support of this research project by the Eliteprogramme for Postdocs.  SY would like to thank the University of T\"ubingen for the financial support and acknowledges the partial support by the Bulgarian NSF Grant KP-06-H28/7. P.N. is partially supported by the Bulgarian NSF Grant  KP-06-H38/2. The authors would also like to acknowledge networking support by the
COST Actions CA16104 and CA15117.

\bibliographystyle{ieeetr}
\bibliography{EGB_axial}

\end{document}